\UseRawInputEncoding
\documentclass[prd,twocolumn,showpacs,amsmath,amssymb]{revtex4}

\usepackage{amssymb}
\usepackage{mathrsfs}
\usepackage{txfonts}
\usepackage{amsmath}

\usepackage{graphicx}
\usepackage{dcolumn}
\usepackage{bm}

\begin{document}

\title{Primordial non-Gaussianity in noncanonical warm inflation with nonminimal derivative coupling}
\author{Xiao-Min Zhang}
\thanks{Corresponding author}
\email{zhangxm@mail.bnu.edu.cn}
\affiliation{School of Science, Qingdao University of Technology, Qingdao 266033, China}
\author{Run-Qing Zhao}
\affiliation{School of Science, Qingdao University of Technology, Qingdao 266033, China}
\author{Yun-Cai Feng}
\affiliation{School of Science, Qingdao University of Technology, Qingdao 266033, China}
\author{Peng-Cheng Chu}
\affiliation{School of Science, Qingdao University of Technology, Qingdao 266033, China}
\author{Zhi-Peng Peng}
\thanks{Corresponding author}
\email{zhipeng@mail.bnu.edu.cn}
\affiliation{School of Physics and Advanced Energy, Henan University of Technology, Zhengzhou 450001, China}
\author{Xi-Bin Li}
\thanks{Corresponding author}
\email{lxbimnu@imnu.bnu.edu.cn}
\affiliation{College of Physics and Electronic Information, Inner Mongolia Normal University, 81 Zhaowuda Road, Hohhot, 010022, Inner Mongolia, China}
\date{\today}

\begin{abstract}
In this paper, the scenario of noncanonical warm inflation with nonminimal derivative coupling is introduced, and its primordial non-Gaussianity of perturbations on uniform energy density hypersurfaces is systematically analyzed. The total non-Gaussianity has two complementary components: the three-point correlation denoted by $f^{int}_{NL}$, and the four-point correlation denoted by $f_{NL}^{\delta N}$. The three-point correlation dominates, and it comes from the intrinsic non-Gaussianity of the inflaton field. In contrast, the four-point correlation is a slow-roll suppressed quantity, which originates from the four-point correlation function of the major Gaussian-part inflaton perturbation and is determined by the background dynamics. These two components of non-Gaussianity are analytically calculated under slow-roll conditions within the noncanonical warm inflation framework with nonminimal derivative coupling. The combined effects of nonminimal derivative coupling, noncanonical kinetic structure, and thermal dissipation crucially determine the non-Gaussian signature, revealing a distinct interplay absent in either cold or minimal warm inflation models. Additionally, comparisons and discussions of these components and the total non-Gaussianity are presented.
\end{abstract}
\maketitle

\section{\label{sec1}Introduction}
The standard model of cosmology has achieved great success in explaining the evolution and observation of the Universe. Nevertheless, it still faces major challenges, including the horizon, flatness, and monopole problems inherent in the Big Bang cosmological framework. Inflation, as a necessary extension to the standard model of cosmology, addresses these issues successfully \cite{Guth1981,Linde1982,Albrecht1982,Bassett2006}. Meanwhile, inflation provides a natural mechanism to explain the observed anisotropy of the cosmic microwave background and the large-scale structures of the Universe. Currently, there are two main types of inflationary theory: standard inflation (generally referred to as cold inflation) and warm inflation. Warm inflation was first proposed by A. Berera in 1995 \cite{BereraFang,Berera1995}. While inheriting the advantages of cold inflation, such as resolving the horizon, flatness, and monopole problems, it introduces distinct features. One of its most prominent features is the relaxation of slow-roll conditions due to thermal dissipative effects \cite{Ian2008,Campo2010,Zhang2014,ZhangZhu}. Warm inflation can effectively resolve the ``$\eta$ problem'' \cite{etaproblem,etaproblem1} and address the issue of an excessively large inflaton amplitude \cite{Berera2006,BereraIanRamos}, both of which pose challenges to cold inflation models. A key difference between cold and warm inflation is the origin of cosmological density fluctuations. In warm inflation, density fluctuations mainly comes from thermal fluctuations \cite{Berera2000,Lisa2004,Taylor2000,Chris2009,BereraIanRamos}, while in cold inflation, they originate from vacuum quantum fluctuations \cite{LiddleLyth,Bassett2006}. During warm inflation, the inflaton field interacts with subdominant boson or fermion fields, resulting in the continuous production of radiation. As a result, a thermal damping term slows down the motion of the inflaton field and eliminates the need for a separate reheating phase. This enables the Universe to smoothly transition into the radiation-dominated phase of the Big Bang.

Traditionally, inflation is realized using a canonical scalar field with the Lagrangian density $\mathcal{L}=X-V$, where $X=\frac12 g^{\mu\nu}\partial_{\mu}\phi\partial_{\nu}\phi$ and $V$ denotes the inflation potential. In addition to this basic framework, various extensions, such as noncanonical fields, pure kinetic fields, multifield setups \cite{refining2012,Mukhanov2006,Armendariz-Picon,Garriga1999,Gwyn2013,Tzirakis,Franche2010,Eassona2013,Bean2008,multifield} and nonminimal coupling fields \cite{Kaiser1995,Karydas2102.08450,GermaniPRL2010}, have been proposed to implement inflation. Over the past three decades, warm inflationary models have also undergone significant developments, including advancements in microphysical foundations \cite{Berera2016,MossXiong}, cosmological perturbations \cite{Berera2000,Chris2009,WangYY2019,Berera2016,Zhang2023,WarmSPy2024,Moss2007}, and extended model frameworks \cite{Zhang2014,Zhang2018,Peng2016,WangYY2018,EadkhongNPB2023}. The most widely investigated warm inflationary models involve a potential-dominated canonical inflaton field within the general relativity (GR) framework. To improve the viability of warm inflationary models, generalized models have been proposed, such as noncanonical warm inflation \cite{Zhang2014,Zhang2018,Zhang2021}, warm $k$-inflation \cite{Peng2016,Peng2018,Zhang2023}, and two-field warm inflation \cite{WangYY2018,WangYY2019}. In these extended warm inflationary models, slow-roll conditions are quite easy to establish compared to the canonical inflationary theory \cite{Zhang2014,Peng2016}. The research on perturbations has confirmed that the inflation energy scale in different warm inflationary models decreases when the horizon crosses, which can be safely described by the effective field theory.

Though inflationary models within the GR framework have been extensively studied, those within modified gravity frameworks mainly focus on cold inflation \cite{Kaiser1995,GermaniPRL2010,EadkhongNPB2023,Karydas2102.08450,DalianisJCAP2020}, and warm inflationary models are rarely explored. In \cite{Sadjadi2015}, the first attempt was made to extend warm inflation to nonminimal gravitational coupling theory, and later, a nonminimal derivative coupling (NMDC) noncanonical warm inflation model was proposed in \cite{Zhang2024}. In the NMDC framework, the inflaton's kinetic term, instead of the inflaton field, is coupled nonminimally to the gravitational sector \cite{Karydas2102.08450,GermaniPRL2010,DalianisJCAP2020,Sadjadi2015}. The NMDC term in this model, $G^{\mu\nu}\partial_{\mu}\phi\partial_{\nu}\phi$ \cite{GermaniPRL2010}, induces enhanced gravitational friction, which slows down the evolution of the scalar field. The presence of nonminimal coupling gravitational friction or thermal damping friction in warm inflation described by $\Gamma\dot\phi$ can make the $\lambda\phi^4$ potential case ruled out in minimal coupling standard inflation again be consistent with the observations \cite{YangNan2015,BeingWarm2014}. The evolution of NMDC noncanonical warm inflationary model is so overdamped to achieve well-relaxed slow-roll conditions, and this model fits the observations well \cite{Zhang2024}. The most attractive feature of the large friction in the evolution equation is owing to the enhanced Hubble damping term, the NMDC gravitational friction term and the thermal damping term, which makes the slow-roll approximations easy to be valid. In addition, the field excursion is good on the sub-Planckian scale.

Existing inflationary research focuses on the power spectrum of density perturbations and the relative amplitude of gravitational waves. However, the power spectrum, as a two-point correlation statistic, provides limited information and fails to distinguish among various inflationary models effectively. Consequently, more and more researchers are interested in higher-order statistics, such as the bispectrum (three-point correlation), trispectrum (four-point correlation), and other measures of deviations from Gaussianity. Non-Gaussianity is a crucial discriminator among inflationary models. The bispectrum represents the lowest-order statistic that can identify non-Gaussian features \cite{Heavens1998,Ferreira1998}. Analyses of non-Gaussianity in canonical \cite{Bartolo2004,Gangui1994,Calzetta1995,DavidLyth2005,Lyth2005,Zaballa2005}, noncanonical and multifield \cite{Creminelli2003,Silverstein2004,Alishahiha2004,Vernizzi2006,Battefeld2007,Tower2010} cold inflation models have indicated that single-field canonical slow-roll inflation produces negligible non-Gaussianity, whereas noncanonical and multifield models can produce significant non-Gaussianity. Warm inflation under the GR framework has also been investigated for non-Gaussianity in various domains, including strong and weak dissipation, as well as temperature-independent and temperature-dependent cases \cite{Moss2007,Gupta2002,Gupta2006,Gil2014}. Nevertheless, analyses of non-Gaussianity in canonical warm inflation have dominated the literature, and limited attention has been given to noncanonical scenarios.

The non-Gaussianity of noncanonical warm inflation was analyzed in \cite{Zhang2015,Zhang2019}, and it was found that both a small sound speed and large dissipation strength improve non-Gaussianity. Our recent studies have investigated primordial non-Gaussianity in warm $k$-inflation \cite{Zhang2023} and two-filed warm inflation \cite{Li2024}, discovering that non-Gaussianity is substantially more obvious compared to potential-dominated slow-roll inflation. In these models, the noncanonical parameter $\mathcal{L}_X$ plays a key role in amplifying non-Gaussianity. Detailed research on the shape of non-Gaussianity in warm inflation \cite{ShapeWI2022} introduced a ``new-warm'' shape, which is different from traditional equilateral templates. Observational constraints are usually expressed in terms of the nonlinear parameter $f_{NL}$, which effectively quantifies the level of non-Gaussianity. In NMDC noncanonical warm inflation, non-Gaussianity has two complementary components: the intrinsic three-point correlation $f^{int}_{NL}$, arising from the inflaton's inherent dynamics, and the four-point correlation $f_{NL}^{\delta N}$, associated with the dominated Gaussian part of the inflaton field. The nonlinear parameter $f_{NL}^{\delta N}$ is evaluated using the $\delta N$ formalism - a well-established approach for analyzing non-Gaussianity, as employed in \cite{DavidLyth2005, Zaballa2005}. In this work, we apply this formalism to compute the four-point correlation within the framework of NMDC noncanonical warm inflation. In particular, although the slow-roll suppression of the $\delta N$ part non-Gaussianity resembles the situation in cold inflation, its realization in the NMDC warm inflation framework is nontrivial. The presence of both the NMDC friction term and the thermal dissipation term relaxes the slow-roll conditions considerably, allowing slow-roll suppression to persist across a much broader parameter space without requiring an extremely flat potential \cite{Zhang2024}. Therefore, the $\delta N$-type component remains small but under a distinct dynamical mechanism compared with standard cold inflation. Hence, the small amplitude of the $\delta N$ non-Gaussianity in this new model necessitates a precise calculation of the intrinsic component. Given the novel features of this model, its non-Gaussianity needs further investigation, which is the focus of this paper.

The rest of this paper is organized as follows. Section \ref{sec2} presents the NMDC noncanonical warm inflationary framework and its governing equations. Section \ref{sec3} introduces the $\delta N$ formalism and evaluates the contributions made by both the four-point correlation and the intrinsic three-point correlation of the inflaton field to non-Gaussianity. Finally, conclusions and discussions are given in Section \ref{sec4}.

\section{\label{sec2}the NMDC noncanonical warm inflationary model}
In warm inflation, a large amount of radiation is produced constantly during the inflationary epoch, and the Universe is hot with a non-zero temperature $T$.
Considering the nonminimally gravitational coupling and interaction between inflaton and sub-dominated fields, the total action for the multi-component warm Universe is represented as \cite{Zhang2024}:
\begin{equation}\label{action}
    S=\int d^4x\sqrt{-g}\left[\frac{1}{2}M_p^2R+\mathcal{L}(X,\phi)+\frac{G^{\mu\nu}}{2M^2}\partial_
    {\mu}\phi\partial_{\nu}\phi+\mathcal{L}_{int}+\mathcal{L}_R\right],
\end{equation}
where $G_{\mu\nu}$ denotes the Einstein tensor, $M_p^2 \equiv {\left( {8\pi G} \right)^{ - 1}}$ stands for the reduced squared Planck mass, and $M$ is the coupling constant with the dimension of mass. It can be seen that the total Lagrangian density contains the gravity, the inflaton field, the additional NMDC $\mathcal{L}_{NMDC}=\frac1{2M^2}G^{\mu\nu} \partial_{\mu}\phi\partial_{\nu}\phi$, the radiation, and the field interactions.
The quantity $X$ is $X=\frac12g^{\mu\nu}\partial_{\mu}\phi\partial_{\nu}\phi$, and $\mathcal{L}(X,\phi)$ represents the Lagrangian density of the noncanonical field, $\mathcal{L}_R$ denotes the Lagrangian density of radiation fields, and $\mathcal{L}_{int}$ stands for the interactions between the scalar fields in warm inflation.

A major evolution equation governing the NMDC noncanonical warm inflation is obtained by varying the total action:
\begin{equation}\label{EOM0}
\frac{\partial(\mathcal{L}(X,\phi)+\mathcal{L}_{int})}{\partial\phi}-\left(\frac{1}{\sqrt{-g}}
\right)\partial_{\mu}\left(\sqrt{-g}\frac{\partial\left(\mathcal{L}(X,\phi)+\mathcal{L}_{NMDC}\right)}
{\partial(\partial_{\mu}\phi)}\right)=0.
\end{equation}
Applying to the flat Friedmann-Robertson-Walker (FRW) Universe, the field is homogeneous, i.e., $\phi=\phi(t)$. Thus, the motion equation for the inflaton field can be reduced to:
\begin{eqnarray}\label{EOM1}
    \left[\left(\frac{\partial\mathcal{L}(X,\phi)}{\partial X}\right)+2X\left(\frac{\partial^2\mathcal{L}(X,\phi)}{\partial X^2}\right)+\frac{3H^2}{M^2}\right]\ddot{\phi} \quad\quad\quad \nonumber \\ +\frac{\partial^2\mathcal{L}(X,\phi)}{\partial X\partial\phi}\dot\phi^2+\left[3H\left(\frac{\partial\mathcal{L}(X,\phi)}{\partial X}+\frac{3H^2}{M^2}+\frac{2\dot H}{M^2}\right)\right]\dot{\phi}\nonumber \\ -\frac{\partial(\mathcal{L}(X,\phi)+
    \mathcal{L}_{int})}{\partial\phi}=0,\quad\quad\quad\quad\quad\quad\quad\quad\quad\quad
\end{eqnarray}
where $X$ is reduced to $X=\frac12\dot\phi^2$, and $H$ is the Hubble parameter. The Hubble parameter is obtained through the Hamiltonian constraint:
\begin{equation}
    3H^{2}=\frac{1}{M_{p}^{2}}(2X \mathcal{L}_{X}-\mathcal{L}+9FX+Ts).
\end{equation}

The sound speed for the noncanonical field characterizes the propagation speed of scalar perturbations, and it is given by $c_{s}^{2}=p_{X}(\phi,X)/\rho_{X}(\phi,X)=\left(1+2X\mathcal{L}_{XX}/\mathcal{L}_{X}\right)^{-1}$, where the subscript $X$ denotes a derivative.

The Lagrangian density of the inflaton for our NMDC noncanonical warm inflation has a separable form, which consists of a kinetic term and a potential term, denoted as $ \mathcal{L}=K(X)-V(\phi)$. Specifically, the noncanonical kinetic term $K$ depends weakly or not at all on $\phi$ \cite{Franche2010}. So, in this scenario, $K$ is assumed to be only a function of $X$, leading to $\mathcal{L}_{X\phi}=0$ and thus $K_{X}=\mathcal{L}_{X}$. The interaction term $\mathcal{L}_{int}$ in Eqs. (\ref{action}) and (\ref{EOM0}) depends on the zeroth order of the inflaton and other fields, but not on their derivatives. To successfully realize warm the inflationary Universe, two types of effective mechanisms, including the supersymmetric two-stage mechanism \cite{BereraKephart,MossXiong} and the warm little inflaton mechanism \cite{Berera2016}, have been developed for the interaction between the inflation and other fields. The term $\Gamma\dot\phi$ can generally describe the dissipation of $\phi$ into other fields \cite{BereraFang,Berera2006,Berera2000,BereraIanRamos}, so it serves as a thermal damping term. The other terms that do not contain $\dot\phi$ in $\partial\mathcal{L}_{int}/\partial\phi$ of Eq. (\ref{EOM1}) and $\partial\mathcal{L}(X,\phi)/\partial\phi$ are introduced into the effective potential $V_{eff}$, which includes thermal corrections and depends on both the inflaton and the temperature. Under these assumptions, the evolution equation for the inflaton is reduced to the following expression \cite{Zhang2024}:
\begin{equation}\label{EOM2}
    (\mathcal{L}_{X}c_{s}^{-2}+3F)\ddot{\phi}+3H\left(\mathcal{L}_{X}+3F+\frac{2\dot{F}}{H}\right)\dot{\phi}
    +\Gamma\dot{\phi}+V_{eff,\phi}=0,
\end{equation}
where the subscripts $X$ and $\phi$ both denote a derivative, and the thermal corrected effective potential $V_{eff}$ is rewritten as $V$ in the following for simplicity. Meanwhile, $F=\frac{H^{2}}{M^{2}}$, and the term $\dot F\ll HF$ during inflationary epoch. Therefore, the term $\frac{2\dot{F}}{H}\dot{\phi}$ in the above equation can be neglected compared to the preceding term. When $F\ll1$, the Einstein GR limit is recovered, and $F\gg1$ indicates a regime of high gravitational friction limit. Additionally, the parameter $\Gamma$ is the dissipative coefficient that describes the thermal damping of the inflaton to radiation, and it can be assumed as a constant or a function of inflaton field and temperature \cite{MossXiong,Berera2016}.

Combining the inflation evolution equation, total energy conservation equation $\dot\rho+3H(\rho+p)=0$ and the total energy density $\rho=2X \mathcal{L}_{X}-K(X)+V(\phi,T)+9FX+Ts$, the entropy production equation can be obtained as:
\begin{equation}\label{entropy1}
    T\dot{s}+3HTs=\Gamma\dot{\phi}^{2}.
\end{equation}
Thermal corrections to the potential are sufficiently small, which holds in the slow-roll regime as Eq. (\ref{SRcondition}) suggested, and then the radiation energy density can be expressed as $\rho_r=3Ts/4$. Thus, Eq. (\ref{entropy1}) is equivalent to the production equation of radiation energy density $\dot{\rho}_r+4H\rho _r=\Gamma \dot{\phi}^2$.

The traditional dissipation strength parameter $r$ in warm inflation is defined as $r=\frac{\Gamma}{3H}$. Considering the comparison between thermal dissipative effect and other non-thermal damping terms, the proper dissipation strength parameter in this NMDC noncanonical warm inflation is represented as
\begin{equation}\label{Q}
 Q=\frac{\Gamma}{3H(\mathcal{L}_{X}+3F)}.
\end{equation}
Then, strong dissipative warm inflation can be characterized by $Q\gg1$, while warm inflation is in the weak regime when $Q\ll1$.

Inflation theory must be predictive, and by neglecting the highest-order terms in Eqs. (\ref{EOM2}) and (\ref{entropy1}), the slow-roll approximation equations are derived:
\begin{equation}\label{EOM3}
    3H(\mathcal{L}_X+3F)\dot{\phi}+\Gamma\dot{\phi}+V_\phi=0,
\end{equation}
\begin{equation}\label{entropy2}
    3HTs=\Gamma\dot{\phi}^{2}.
\end{equation}
Compared to the minimal coupling case in cold canonical inflation, the damping term is enhanced by a factor of ($\mathcal{L}_{X}+3F+r$), suggesting that NMDC noncanonical warm inflation is strongly overdamped and thus the slow roll occurs naturally and necessarily.

The slow-roll parameters in warm inflationary pictures are usually defined as:
\begin{equation}
 \epsilon =\frac{M_p^2}{2}\left(\frac{V_{\phi}}{V}\right) ^2,~~\eta =M_p^2\frac {V_{\phi \phi}}{V},~~ \beta
=M_p^2\frac{V_{\phi}\Gamma_{\phi}}{V\Gamma},
\end{equation}
To describe the temperature dependence in warm inflation, another two slow-roll parameters are given by:
\begin{equation}
 b=\frac {TV_{\phi T}}{V_{\phi}},~~~ c=\frac{T\Gamma_T}{\Gamma}.
\end{equation}

The validity of slow-roll approximations is guaranteed by the slow-roll conditions \cite{Zhang2024}:
\begin{equation}\label{SRcondition}
    \begin{aligned}\epsilon<\mathcal{L}_{X}+3F+r, \quad \beta<\frac{(\mathcal{L}_{X}c_s^{-2}+3F)(\mathcal{L}_{X}+3F+r)}{r},\\ \eta<\mathcal{L}_{X}c_s^{-2}+3F,\quad b<\frac{\mathcal{L}_{X}c_s^{-2}+3F}{\mathcal{L}_{X}+3F+r}, \quad  |c|<4.\end{aligned}
\end{equation}
Compared to cold and warm inflation in the GR frame, the slow-roll conditions in this scenario are much more relaxed, making it very easy to establish the slow-roll approximations regardless of the concrete form of the scalar potential.

The number of e-folds in NMDC noncanonical warm inflation is given by
\begin{equation}
N=\int H dt=\int\frac{H}{\dot{\phi}}d\phi\simeq-\frac{1}{M_p^2}\int_{\phi_{\ast}}
    ^{\phi_{end}}\frac{V(\mathcal{L}_X+3F+r)}{V_{\phi}}d\phi,
\end{equation}
where $\phi_{\ast}$ denotes the field value at the crossing of the Hubble horizon. Due to the overdamped feature and easily satisfied slow-roll conditions of this scenario, the number of e-folds is sufficient to resolve the three major problems of the standard inflationary model.

\section{\label{sec3}the non-Gaussianity in NMDC noncanonical warm inflation}

In the spatially flat gauge, the curvature perturbation on uniform density hypersurfaces is defined as $\zeta=\frac{H}{\dot\rho}\delta\rho$ \cite{Bassett2006}. In the single-field warm inflationary scenario, the relation $\zeta=\frac{H}{\dot\phi}\delta\phi$ still holds, and the curvature perturbation $\zeta$ is still conserved on large scales once it becomes superhorizon \cite{PhDBerera2021,Das2020}. The power spectrum of the curvature perturbation, denoted by $P_{\zeta}$, is defined through the following relation:
\begin{equation}\label{spectrum}
  \langle\zeta_{\mathbf{k}_1}\zeta_{\mathbf{k}_2}\rangle\equiv(2\pi)^3\delta^3(\mathbf{k}_1+\mathbf{k}_2)P_{\zeta}(k_1),
\end{equation}
and the commonly used dimensionless scale-free power spectrum $\mathcal{P}_{\zeta}(k)$ is related to $P_{\zeta}$ as $\mathcal{P}_{\zeta}(k)\equiv\frac{k^3}{2\pi^2}P_{\zeta}(k)$.

The three-point function of the curvature perturbation, or its Fourier transform (the bispectrum), is the main and lowest-order measure of non-Gaussianity. The bispectrum is defined through the following relation:
\begin{equation}\label{bispectrum0}
  \langle\zeta_{\mathbf{k}_1}\zeta_{\mathbf{k}_2}\zeta_{\mathbf{k}_3}\rangle\equiv(2\pi)^3\delta^3
(\mathbf{k}_1+\mathbf{k}_2+\mathbf{k}_3)B_{\zeta}(k_1,k_2,k_3).
\end{equation}

Observational limits are usually represented in terms of the nonlinear parameter $f_{NL}$, which quantifies the magnitude of non-Gaussianity. For curvature perturbations with an approximate Gaussian distribution, the bispectrum has a general form:
\begin{equation}\label{Bzeta}
  B_{\zeta}(k_1,k_2,k_3)\equiv -\frac65f_{NL}(k_1,k_2,k_3)\left[P_{\zeta}(k_1)P_{\zeta}(k_2)+cyclic \right].
\end{equation}

The non-Gaussianity in the NMDC noncanonical warm inflation model consists of two complementary components: the $\delta N$ part and the intrinsic part. The total non-Gaussianity is described as $f_{NL}=f_{NL}^{\delta N}+f^{int}_{NL}$, and our analysis is based on this relation. In the following, the two components are calculated separately.

\subsection{\label{sec31}the $\delta N$-part non-Gaussianity}

The $\delta N$ formalism, commonly used to calculate the non-Gaussianity in multifield inflation \cite{Vernizzi2006,Battefeld2007,Tower2010}, has been extensively developed in previous studies \cite{Lyth2005,DavidLyth2005,Starobinsky,Sasaki1996,Sasaki1998}. Cosmological observations indicate that the primordial curvature perturbation $\zeta$ on uniform density hypersurfaces is dominated by Gaussian terms with a nearly scale-invariant spectrum.

During the inflationary epoch, the evolution of the Universe is assumed to be mainly governed by one or more inflaton fields. Including small perturbations, the full scalar field can be represented as $\Phi_i(t,\mathbf{x})=\phi_i(t)+\delta\phi_i(t,\mathbf{x})$ in a convenient flat-slicing gauge. Observations suggest that $\zeta$ is nearly Gaussian, making it possible to expand $\zeta$ up to second order for sufficient accuracy:
\begin{equation}\label{zeta2}
  \zeta(t,\mathbf{x})=\delta N\simeq \sum_i N_{,i}(t)\delta\phi_i+\frac12\sum_{ij} N_{,ij}(t)\delta\phi_i\delta\phi_j,
\end{equation}
where $N_{,i}\equiv\frac{\partial N}{\partial\phi_i}$ and $N_{,ij}\equiv\frac{\partial^2 N}{\partial\phi_i\partial\phi_j}$.

If the field perturbations are purely Gaussian, this expansion indicates that the four-point correlation contributes entirely to non-Gaussianity, and it is referred to as the $\delta N$-part non-Gaussianity. However, inflaton field perturbations in various inflationary models deviate from pure Gaussian distributions. Such deviations are more obvious in noncanonical inflationary scenarios than in canonical ones. As a result, it is also crucial to calculate non-Gaussianity generated by the intrinsic interaction mechanism of the inflaton field.

Using Eq. (\ref{zeta2}), the total non-Gaussianity can be calculate as:
\begin{eqnarray}\label{threepoint}
  \langle\zeta_{\mathbf{k}_1}\zeta_{\mathbf{k}_2}\zeta_{\mathbf{k}_3}\rangle =\sum_{ijk}N_{,i}N_{,j}N_{,k}
  \langle\delta\phi^{i}_{\mathbf{k}1}\delta\phi^{j}_{\mathbf{k}2}\delta\phi^{k}_{\mathbf{k}3}\rangle ~~~~~~~~~~~\nonumber\\
  +\frac12\sum_{ijkl}N_{,i}N_{,j}N_{,kl}\langle\delta\phi^{i}_{\mathbf{k}1}\delta\phi^{j}_{\mathbf{k}2}
  (\delta\phi^{k}\star\delta\phi^{l})_{\mathbf{k}3}\rangle+perms,
\end{eqnarray}
where, the star denotes convolution, and contributions from higher-order correlation functions beyond the four-point part are negligible. The first term denotes intrinsic non-Gaussianity from the inflaton field, while the second term corresponds to $\delta N$-part non-Gaussianity, which is scale-independent and can be easily calculated using the $\delta N$ formalism.

Using the $\delta N$ formalism, the nonlinear parameter for the $\delta N$ part is given by \cite{Lyth2005,Boubekeur}:
\begin{equation}\label{fNL}
  -\frac35 f_{NL}^{\delta N}= \frac{\sum\limits_{ij} N_{,i}N_{,j}N_{,ij}}{2\left[\sum\limits_{i} N^2_{,i}\right]^2}.
\end{equation}

In NMDC noncanonical warm inflation, a single inflation field dominates, and only one $\delta\phi_i$ is relevant, so Eq. (\ref{zeta2}) reduces to:
\begin{equation}\label{zeta3}
  \zeta(t,\mathbf{x})=N_{,i}\delta\phi_i+\frac12 N_{,ii}\left(\delta\phi_i\right)^2,
\end{equation}
This simplifies
\begin{equation}\label{fNL1}
  -\frac35f_{NL}^{\delta N}=\frac12\frac{N_{,ii}}{N^2_{,i}}.
\end{equation}
Since there is only one $\delta\phi_i$, without ambiguity, $N_{,i}$ as $N_{\phi}$ and $N_{,ii}$ can be rewritten as $N_{\phi\phi}$ below. Thus, the relation reduces to $-\frac35f_{NL}^{\delta N}=\frac12\frac{N_{\phi\phi}}{N_\phi^2}$.
The $f_{NL}^{\delta N}$ term is scale-independent, which can be found in Eq. (\ref{fNL1}).

At horizon crossing, the nonlinear parameter $f_{NL}^{\delta N}$ of the $\delta N$ part under slow-roll approximations can be calculated as:
\begin{equation}\label{Nphi}
  N_\phi=-\frac1{M_p^2}\frac{V(\mathcal{L}_X+3F+r)}{V_\phi}.
\end{equation}

From Eq. (\ref{Nphi}), the second derivative is:
\begin{eqnarray}\label{Nphiphi}
 N_{\phi\phi}&=&-\frac{1}{M_{p}^{2}}\Bigg[(\mathcal{L}_{X}+3F+r)-\frac{VV_{\phi\phi}(\mathcal{L}_{X}+3F+r)}{V_{\phi}^{2}}\nonumber \\&+&\frac{V}{V_{\phi}}\frac{\Gamma_{\phi}}{3H}-\frac{r}{2}+3F\Bigg].
\end{eqnarray}
And then we have
\begin{eqnarray}\label{NphiNphiphi}
  \frac{N_{\phi\phi}}{N_{\phi}^2}=-M_p^2\left[\frac{V_{\phi}^2}{(\mathcal{L}_X+3F+r)V^2}+\frac{V_{\phi\phi}}
  {(\mathcal{L}_X+3F+r)V}\right.\nonumber\\ \left. +\frac{\Gamma_{\phi}V_{\phi}r}{\Gamma V(\mathcal{L}_X+3F+r)^2}-\frac{rV_{\phi}^2}{2V^2(\mathcal{L}_X+3F+r)^2}\right.\nonumber\\ \left.+\frac{3FV^2_{\phi}}{V^2(\mathcal{L}_X+3F+r)^2}\right].\quad\quad\quad\quad\quad\quad\quad\quad
\end{eqnarray}

With the assistance of Eqs. (\ref{fNL1}) and (\ref{NphiNphiphi}) and the slow-roll parameters $\epsilon=\frac{M_p^2}{2}\left(\frac{V_{\phi}}{V}\right) ^2$, $\eta=M_p^2\frac {V_{\phi \phi}}{V}$, $\beta=M_p^2\frac{V_{\phi}\Gamma_{\phi}}{V\Gamma}$, the nonlinear parameter for $\delta N$ part can be represented as:
\begin{eqnarray}\label{fNLdeltaN}
  f_{NL}^{\delta N}=-\left[\frac{2\epsilon}{\mathcal{L}_X+3F+r}+\frac\eta{\mathcal{L}_X+3F+r}\right.\nonumber\\
  \left.+\frac{r\beta}{\left(\mathcal{L}_X+3F+r\right)^2}+\frac{(6F-r)\epsilon}{\left(\mathcal{L}_X+3F+r\right)^2}\right].
\end{eqnarray}

From these calculations, it can be seen that the $\delta N$-part non-Gaussianity is scale-independent because it is determined only by non-perturbative background equations, as indicated by the $\delta N$ formalism. Under slow-roll conditions in NMDC noncanonical warm inflation, $|f_{NL}^{\delta N}|\sim \mathcal{O}\left(\frac{\epsilon}{\mathcal{L}_{X}+3F+r}\right)<1$, suggesting that $f_{NL}^{\delta N}$ is slow-roll suppressed, as expected. And this is consistent with other warm inflationary scenarios \cite{Zhang2015,Zhang2016}. However, its physical origin in the NMDC warm inflation framework differs from the standard inflation cases. The suppression arises not solely from the smallness of the slow roll parameters, but from the combined frictional effects of the NMDC term and the thermal dissipation, which jointly maintain the overdamped evolution of the inflaton field. This implies that the $\delta N$ contribution remains subdominant even in parameter regions where the potential is moderately steep, highlighting a dynamical robustness not present in canonical inflation.

If the field perturbations are purely Gaussian, the $\delta N$-part non-Gaussianity can represent the total primordial non-Gaussianity generated by inflation even though it is relatively weak during the inflationary epoch. Since the deviation from pure Gaussian distribution is enlarged in this NMDC noncanonical warm inflation, the $\delta N$ part contribution is insufficient to explain the total primordial non-Gaussianity \cite{DavidLyth2005,Zhang2016}. Therefore, it is necessary to calculate the part of non-Gaussianity arising from inflaton's intrinsic interaction mechanisms.

\subsection{\label{sec32}the intrinsic part of non-Gaussianity}
Unlike cold inflation, the perturbations in warm inflation mainly originate from thermal fluctuations. Consequently, the way to calculate cosmological perturbations differs from that used in cold inflation.
In a warm inflationary multi-component Universe, only one scalar field acts as the inflaton. Considering small perturbations, the full inflaton field can be expanded as $\Phi(\mathbf{x},t)=\phi(t)+\delta\phi(\mathbf{x},t)$, where $\delta\phi(\mathbf{x},t)$ denotes the perturbation field around the homogeneous background, which is a linear response to the thermal stochastic noise $\xi$ in the thermal system \cite{Berera2000,BereraIanRamos}. The thermal stochastic noise $\xi$ in the thermal system has zero mean $\langle\xi\rangle=0$, and under the high-temperature limit $T\rightarrow\infty$, the noise source becomes Markovian: $\langle\xi(\mathbf{k},t)\xi(-\mathbf{k'},t')\rangle=2\Gamma Ta^{-3}(2\pi)^3\delta^3(\mathbf{k}-\mathbf{k'})\delta(t-t')$ \cite{Lisa2004,Gleiser1994}. By introducing the thermal white noise term and deriving from the full inflaton field equation Eq. (\ref{EOM0}), the second-order Langevin equation for the perturbation can be obtained \cite{Zhang2024}, but the process of solving it is quite complex:
\begin{widetext}
\begin{equation}\label{pEOM1}
    \begin{aligned}(\mathcal{L}_{X}c_{s}^{-2}+3F)(\ddot{\phi}(t)+\delta\ddot{\phi}(\mathbf{x},t))+
    \left[3H(\mathcal{L}_{X}+3F)+\Gamma\right]
    (\dot{\phi}(t)+\delta\dot{\phi}(\mathbf{x},t))\\+V_{\phi}+V_{\phi\phi}\delta\phi(\mathbf{x},t)-
    \mathcal{L}_{X}\frac{\nabla^{2}}{a^{2}}\delta\phi(\mathbf{x},t)+3Fw\frac{\nabla^{2}}{a^{2}}\delta\phi(\mathbf{x},t)
    =\xi(\mathbf{x},t).\end{aligned}
\end{equation}

Perturbation quantities are evaluated at the time of Hubble horizon crossing, which occurs in the slow-roll regime. In the slow-roll regime of NMDC noncanonical warm inflation, due to the greatly enhanced damping term, the evolution of the inflation is overdamped. Considering this, the inertia term in the evolution of the inflaton perturbations can be neglected, as indicated by \cite{Berera2000,Zhang2024}. In this regime, the equation of state simplifies to $w=\frac{p}{\rho}\simeq-1$. By applying the Fourier transform, the evolution equation for the perturbation field in Fourier space is given by:
\begin{equation}\label{pEOM2}
[3H(\mathcal{L}_{X}+3F)+\Gamma]\delta\dot{\phi_{\mathbf{k}}}+\left(\mathcal{L}_{X}k^2+3Fk^{2}
+m^2\right)\delta\phi_{\mathbf{k}}=\xi_{\mathbf{k}},
 \end{equation}
where, $m^2=V_{\phi\phi}$ stands for the effective squared inflaton mass, $\mathbf{k}$ denotes the physical momentum, with $\mathbf{k}\equiv\mathbf{k_p}=\mathbf{k_c}/a$, and $\mathbf{k}_c$ denotes the comoving momentum. The magnitude of the physical momentum is represented as $k=|\mathbf{k}|$.

The thermal noise term $\xi$ introduced in warm inflation is white noise, indicating that it is Gaussian-distributed \cite{Berera2000}. The inflaton perturbations can be divided into first-order, second-order and higher-order ones (higher-order perturbations are small and thus not considered here). The first-order perturbation is the leading-order inflaton perturbation, and it is also Gaussian-distributed because it is the linear response to the thermal noise. To calculate the bispectrum of the inflaton perturbation, the perturbation must be expanded at least to the second order. The inflaton perturbation is expanded as $\delta\phi(\mathbf{x},t)=\delta\phi_1(\mathbf{x},t)+\delta\phi_2(\mathbf{x},t)$, where $\delta\phi_1=\mathcal{O}(\delta\phi)$ and $\delta\phi_2=\mathcal{O}(\delta\phi^2)$.

The equations of motion for the first and second-order perturbations can be derived by perturbing the evolution equation of the inflaton perturbation Eq. (\ref{pEOM2}) to second order:
\begin{equation}\label{deltaphi1}
 \frac{d}{dt}\delta\phi_{1}(\mathbf{k},t)=\frac{1}{3H(\mathcal{L}_X+3F+r)}[-V_{\phi\phi}\delta\phi_{1}(\mathbf{k},t)-(\mathcal{L}_X+3F)k^{2}\delta\phi_{1}(\mathbf{k},t)+\xi(\mathbf{k},t)],
\end{equation}
and
\begin{eqnarray}\label{deltaphi2}
    \frac{d}{dt}\delta\phi_2(\mathbf{k},t)=\frac{1}{3H(\mathcal{L}_X+3F+r)}\left\{-\left[(\mathcal{L}_X+3F)k^2+V_{\phi\phi}\right]\delta\phi_2\big(\mathbf{k},t\big)\right.
    \quad\quad\quad\quad\quad\quad\quad\quad\quad\quad \nonumber\\
    \left.-\left(\frac{1}{2}V_{\phi\phi\phi}+3k^2F_\phi\right)\int\frac{d^3p}{\left(2\pi\right)^3}\delta\phi_1(\mathbf{p},t)\delta\phi_1(\mathbf{k}-\mathbf{p},t)
    -\left(k^2\mathcal{L}_{XX}+3k^2F_X\right)\int\frac{d^3p}{\left(2\pi\right)^3}\delta\phi_1(\mathbf{p},t)\delta X_1(\mathbf{k}-\mathbf{p},t)\right\}.
\end{eqnarray}
where $\delta X_1 =\dot{\phi}\dot{\delta \phi_1} = \sqrt{2X} \frac{d}{dt} \delta \phi_1$ in above equation.

By solving these two evolution equations, the analytic solutions for the first-order perturbation $\delta\phi_1$ and second-order perturbation $\delta\phi_2$ can be obtained as:
\begin{eqnarray}\label{solution1}
    \delta\phi_1(\mathbf{k},t)=\frac1{3H(\mathcal{L}_X+3F+r)}\mathrm{exp}\left[-\frac{(\mathcal{L}_X+3F)k^2+m^2}{3H(\mathcal{L}_X+3F+r)}(t-t_0)\right]
    \int_{t_0}^t\exp\left[\frac{(\mathcal{L}_X+3F)k^2+m^2}{3H(\mathcal{L}_X+3F+r)}\Big(t^{\prime}-t_0\Big)\xi\Big(\mathbf{k},t^{\prime}\Big)dt^{\prime}\right]\nonumber\\
    +\delta\phi_1(\mathbf{k},t_0)\mathrm{exp}\left[-\frac{(\mathcal{L}_X+3F)k^2+m^2}{3H(\mathcal{L}_X+3F+r)}(t-t_0)\right], \quad\quad\quad\quad\quad\quad\quad\quad\quad\quad\quad \quad\quad\quad\quad
\end{eqnarray}
and
\begin{eqnarray}\label{solution2}
    \delta\phi_2(\mathbf{k},t)=\exp\left[-\frac{(\mathcal{L}_X+3F)k^2+m^2}{3H(\mathcal{L}_X+3F+r)}(t-t_0)\right]
    \int_{t_0}^tdt^{\prime}\exp\left[\frac{(\mathcal{L}_X+3F)k^2+m^2}{3H(\mathcal{L}_X+3F+r)}\left(t^{\prime}-t_0\right)\right]
    \left[A\left(k,t^{\prime}\right)\int\frac{d^3p}{\left(2\pi\right)^3}\delta\phi_1\left(\mathbf{p},t^{\prime}\right)\delta\phi_1\left(\mathbf{k}-\mathbf{p},t^{\prime}\right)\right.\nonumber\\
    \left.-B\left(k,t^{\prime}\right)\int\frac{d^3p}{\left(2\pi\right)^3}\delta\phi_1\left(\mathbf{p},t^{\prime}\right)\xi\left(\mathbf{k}-\mathbf{p},t^{\prime}\right)\right]
    +\delta\phi_2(\mathbf{k},t_0)\mathrm{exp}\left[-\frac{(\mathcal{L}_X+3F)k^2+m^2}{3H(\mathcal{L}_X+3F+r)}(t-t_0)\right], \quad\quad\quad\quad\quad\quad\quad\quad\quad\quad
\end{eqnarray}
\end{widetext}
where the parameters $A(k,t)$ and $B(k,t)$ are given by
\begin{eqnarray}\label{A}
  A(k,t)=\sqrt{2X}(\mathcal{L}_{XX}+3F_{X})k^{2}\frac{(\mathcal{L}_{X}+3F)k^{2}+m^{2}}{\left[3H(\mathcal{L}_{X}+3F+r)\right]^{2}}\nonumber\\
  -\frac{V_{\phi\phi\phi}+6k^2F_\phi}{6H(\mathcal{L}_X+3F+r)},\quad\quad\quad\quad\quad\quad\quad\quad
\end{eqnarray}
and
\begin{equation}
    B(k,t)=\frac{\sqrt{2X}\left(k^2\mathcal{L}_{XX}+3k^2F_X\right)}{\left[3H(\mathcal{L}_X+3F+r)\right]^2}.
\end{equation}
Both the second terms on the right-hand side of the above solutions represent ¡°memory¡± terms, reflecting the state of the mode at the initial time $\tau_0$ in a Hubble time interval \cite{Berera2000}.
From Eq. (\ref{solution1}), it can be seen that the larger the squared magnitude of the physical momentum $k^2$, the faster the relaxation. If $k^2$ is sufficiently large for the mode to relax within a Hubble time interval, the mode thermalizes. As soon as the physical momentum of a $\Phi(\mathbf{x},t)$ field mode decreases below a critical value $k_F$, it will experience no thermal noise during a Hubble time interval. The freeze-out momentum $k_F$ in warm inflation is defined as the critical value at which the mode ceases to experience the thermal noise. Based on Eq. (\ref{solution1}), a parameter $\tau(\phi)=\frac{3H(\mathcal{L}_X+3F+r)}{(\mathcal{L}_X+3F)k^2+m^2}$ is thus defined to describe the efficiency of the thermalizing process. The thermalizing criterion above indicates that the freeze-out physical momentum $k_F$ can be defined by $\tau(\phi)=\frac{1}{H}$. The mass term can be ignored under the slow-roll conditions, and the freeze-out momentum $k_F$ is then represented as
\begin{eqnarray}\label{kf}
   k_F&=&H\sqrt{\frac{3(\mathcal{L}_X+3F+r)}{\mathcal{L}_X+3F}}\nonumber\\&=&H\sqrt{3(1+Q)}.
\end{eqnarray}

As it is described earlier, the first-order inflaton perturbations $\delta\phi_1$ are Gaussian fields. Stochastic statistical properties indicate that their bispectrum vanishes. To calculate non-Gaussianity, the leading contribution originates from the bispectrum generated by two first-order and one second-order fluctuations:
\begin{widetext}
\begin{eqnarray}\label{threepoint1}
\langle\delta\phi(\mathbf{k}_1,t)\delta\phi(\mathbf{k}_2,t)\delta\phi(\mathbf{k}_3,t)\rangle \quad=\quad \exp\left[-\frac{(\mathcal{L}_X+3F)k^2+m^2}{3H(\mathcal{L}_X+3F+r)}(t-t_0)\right]\int_{t_0}^tdt^{\prime}\exp\left[\frac{(\mathcal{L}_X+3F)k^2+m^2}{3H(\mathcal{L}_X+3F+r)}\left(t^{\prime}
    -t_0\right)\right] \quad\quad\quad\quad\quad\quad\quad \nonumber\\
    \left[A\left(k_3,t^{\prime}\right)\int\frac{d^3p}{\left(2\pi\right)^3}\langle\delta\phi_1(\mathbf{k}_1,t)\delta\phi_1(\mathbf{k}_2,t)\delta\phi_1\left(\mathbf{p},t^{\prime}\right)
    \delta\phi_1\left(\mathbf{k}_3-\mathbf{p},t^{\prime}\right)\rangle
    -B\left(k_3,t^{\prime}\right)\int\frac{d^3p}{\left(2\pi\right)^3}\langle\delta\phi_1(\mathbf{k}_1,t)\delta\phi_1(\mathbf{k}_2,t)\delta\phi_1\left(\mathbf{p},t^{\prime}\right)\xi
    \left(\mathbf{k}_3-\mathbf{p},t^{\prime}\right)\rangle\right]\nonumber\\
    +\exp\left[-\frac{(\mathcal{L}_X+3F)k^2+m^2}{3H(\mathcal{L}_X+3F+r)}(t-t_0)\right]\langle\delta\phi_1(\mathbf{k}_1,t)\delta\phi_1(\mathbf{k}_2,t)\delta\phi_2(\mathbf{k}_3,t_0)\rangle
    +\left(\mathbf{k}_1\leftrightarrow \mathbf{k}_3\right)+\left(\mathbf{k}_2\leftrightarrow \mathbf{k}_3\right). \quad\quad\quad\quad\quad\quad\quad\quad\quad\quad\quad\quad
\end{eqnarray}
\end{widetext}
The amplitude of the bispectrum and the power spectrum is determined when cosmological scales exit the horizon, and the time is approximately 60 e-folds before the end of inflation, with $\mathbf{k}_1$, $\mathbf{k}_2$ and $\mathbf{k}_3$ all within a few e-folds of horizon crossing. The non-Gaussian amplitude peaks in the shape of nearly equilateral triangles, with $k_1\approx k_2\approx k_3$, and it vanishes in the shape of squeezed triangles, where one scale is much smaller than the others, i.e., $k_1\ll k_2,k_3$. Therefore, the intrinsic non-Gaussianity in this NMDC noncanonical warm inflation model is in a nearly equilateral shape.

From the expression of $k_F$, we have $k_F>H$ in the NMDC noncanonical warm inflationary model. This suggests that the $\delta\phi$ correlations, which should be calculate at the time of Hubble horizon crossing $k=H$, are actually thermalized correlations that were fixed at an earlier freeze-out time $k=k_F$ \cite{Berera2000}. As a result, the time interval in the corrections can be given by:
\begin{equation}\label{time}
\Delta t_F=t_H-t_F\simeq\frac{1}{H}\ln\left(\frac{k_F}{H}\right).
\end{equation}
As a result, the bispectrum reduces to:
\begin{eqnarray}\label{threepoint2}
 \langle\delta\phi(\mathbf{k}_1,t)\delta\phi(\mathbf{k}_2,t)\delta\phi(\mathbf{k}_3,t)\rangle  \quad\quad\quad\quad\quad\quad\quad \quad\quad\quad\nonumber\\
    \simeq2A(k_F,t)\Delta t_F\left[1-\frac{2r}{(\mathcal{L}_X+3F)(\mathcal{L}_X+3F+r)}\right] \quad\quad\quad\nonumber\\
    \left[\int\frac{d^3p}{\left(2\pi\right)^3}\langle\delta\phi_1(\mathbf{k}_1,t)\delta\phi_1(\mathbf{p},t)\rangle\langle\delta\phi_1(\mathbf{k}_2,t)\delta\phi_1
    (\mathbf{k}_3-\mathbf{p},t)\rangle\right.\nonumber\\
    \left.+\left(\mathbf{k}_1\leftrightarrow\mathbf{k}_3\right)+\left(\mathbf{k}_2\leftrightarrow\mathbf{k}_3\right)\right], \quad\quad\quad\quad\quad\quad\quad\quad\quad
\end{eqnarray}
by considering the statistical properties of the inflaton perturbations and thermal noise.

In the spatially flat gauge, the relation $\zeta=\frac{H}{\dot\phi}\delta\phi$ holds as referenced above.
Using Eqs. (\ref{threepoint2}) and (\ref{Bzeta}), the nonlinear parameter of intrinsic non-Gaussianity can be obtained as:
\begin{eqnarray}\label{fNLint}
f_{NL}^{int}\simeq-\frac{5}{6}\frac{\dot{\phi}}{H}2A(k_F,t_F)\Delta t_F\nonumber \quad\quad\quad\quad\quad\quad\quad\quad\quad\quad\\
    =\mathrm{ln}\frac{k_F}{H}\left\{-\frac{5}{3}\left[\frac{\mathcal{L}_{X}}{\mathcal{L}_{X}+3F}(c_{s}^{-2}-1)+2Z\frac{\mathcal{L}_{X}c_{s}^{-2}+9F}{\mathcal{L}_{X}+3F}\right]\right.\nonumber \\
    \left.-\frac{5}{6}\frac{\epsilon\varepsilon}{\left(\mathcal{L}_X+3F+r\right)^2}-\frac{10F\epsilon}{(\mathcal{L}_X+3F)(\mathcal{L}_X+3F+r)}\right\},
\end{eqnarray}
where the parameter arising from gravitationally nonminimal coupling is $Z=\frac{X}{M_p^2M^2}=\frac{3FX}{\rho}<1$, which measures the relative importance of the nonminimal derivative coupling within the multi-component inflationary system, and the parameter $\varepsilon$ is defined as $\varepsilon=2M_p^2\frac{V_{\phi\phi\phi}}{V_\phi}$, which can be regarded as a small slow-roll quantity similar to the slow-roll parameter $\eta$ in the case of a monomial potential. Therefore, the intrinsic nonlinear parameter $f_{NL}^{int}$ in the NMDC noncanonical warm inflationary theory is obtained, and it can be compared with observations.
Now, the full expression of $f_{NL}^{int}$ can be rewritten with two terms:
\begin{widetext}
\begin{eqnarray}
f_{NL}^{int}=\underbrace{-\frac{5}{3}\left[\frac{\mathcal{L}_X}{\mathcal{L}_X+3F}(c_s^{-2}-1)+2Z\frac{\mathcal{L}_Xc_s^{-2}+9F}{\mathcal{L}_X+3F}\right]
\ln\sqrt{\frac{3(\mathcal{L}_X+3F+r)}{\mathcal{L}_X+3F}}}_{term1}\nonumber\\
\underbrace{-\left[\frac{5}{6}\frac{\epsilon\varepsilon}{\left(\mathcal{L}_X+3F+r\right)^2}+\frac{10F\epsilon}{\left(\mathcal{L}_X+3F\right)\left(\mathcal{L}_X+3F+r\right)}
\right]\ln\sqrt{\frac{3(\mathcal{L}_X+3F+r)}{\mathcal{L}_X+3F}}}_{term2}.
\end{eqnarray}
\end{widetext}
Applying the slow-roll conditions in NMDC noncanonical warm inflation, we have $term2\ll1$.
Obviously, the first term of the intrinsic nonlinear parameter $f_{NL}^{int}$ is much larger than the second term, so the first term dominates in $f_{NL}^{int}$. Therefore, the intrinsic nonlinear parameter can be approximated as:
\begin{equation}\label{intrinsic}
  f_{NL}^{int}\simeq-\frac{5}{3}\left[\frac{\mathcal{L}_X}{\mathcal{L}_X+3F}(c_s^{-2}-1)+2Z\frac{\mathcal{L}_Xc_s^{-2}+9F}{\mathcal{L}_X+3F}\right]
\ln\sqrt{3(1+Q)}.
\end{equation}
The $Z$-term comes purely from the nonminimal derivative coupling, and the other terms in the above expression for $f_{NL}^{int}$ are also impacted by the NMDC effect. The NMDC term introduces an additional gravitational friction that, together with thermal damping, modifies the inflaton dynamics and slows down its evolution. This enhanced friction partially counteracts the non-Gaussian amplification caused by the noncanonical kinetic term, leading to a moderated $c_s^{-2}$ enhancement compared to the cold-inflation limit. This interplay is analytically reflected in the first term of above equation. The NMDC also yields a distinct non-Gaussian contribution scaled by the characteristic parameter $Z=\frac{3FX}{\rho}$, i.e. the second term in above equation. This component dominates in the strong-coupling regime ($F \gg 1$), while the conventional $c_s^{-2}$ term prevails for weak coupling ($F \ll 1$). This dual structure of $f_{NL}$ is unique to the NMDC noncanonical warm inflation framework and has not been reported in earlier studies of noncanonical or warm inflation in general relativity frame. In addition, the thermal effect significantly affects the non-Gaussianity, particularly in the strong regime where $Q\gg1$. In the very weak dissipative regime, where $Q\ll1$, the non-Gaussian result is more influenced by the NMDC and noncanonical effects. The features of novel synergy among nonminimal derivative coupling, noncanonical kinetic effects, and thermal dissipation in shaping the non-Gaussian signatures are absent in both GR-based warm inflation and cold noncanonical models.

In contrast to inflation with a canonical kinetic term, the non-Gaussian features in this case demonstrate an obvious $(c_s^{-2}-1)$ dependence, as in noncanonical inflationary scenarios. Based on the result in Eq. (\ref{intrinsic}), a small sound speed of inflaton can substantially improve the level of intrinsic non-Gaussianity. This suggests that the stronger the noncanonical effect, the larger the deviation from a Gaussian distribution. The quantities $F$ and $Z$ also play a crucial role in determining the level of non-Gaussianity. And from the result, it can be seen that when the $F$ value is not large, the noncanonical effect dominates the non-Gaussianity, and when $F$ becomes sufficiently large, the nonminimal coupling effect becomes more prominent. Meanwhile, it is found that the strong dissipation effect of warm inflation contributes to intrinsic non-Gaussianity to some extent. Compared to single-field slow-roll cold inflation, NMDC noncanonical warm inflation exhibits a greater degree of non-Gaussianity, due to the mutual enhancement of the nonminimal derivative coupling, noncanonical effects, and thermal damping. A significant non-Gaussian level will favor this model, and these three effects should be very weak if observations exhibit no significant primordial non-Gaussianity \cite{PLANCKNG2018}.

\subsection{\label{sec33}estimation of the total non-Gaussianity}

Based on the results obtained in the preceding section, the non-Gaussianities in two parts are now compared. As previously described, $f_{NL}^{\delta N}$ can be represented as a polynomial function of the slow-roll parameters. As a result, $f_{NL}^{\delta N}$ cannot be much larger than one. Meanwhile, the intrinsic part $f^{int}_{NL}$ can be much larger than one if the noncanonical, thermal, and NMDC effects are sufficiently strong. From this, it can be concluded that the dominant contribution to the total non-Gaussianity comes from the intrinsic part. In this NMDC noncanonical warm inflationary model, the small sound speed and the large NMDC term mainly contribute to the non-Gaussian distributions.

By combining the nonlinear parameters from the two parts, the total nonlinear parameter can be approximated as:
\begin{widetext}
\begin{eqnarray}\label{fNLtotal}
  f^{total}_{NL}&=&f_{NL}^{\delta N}+f_{NL}^{int}\nonumber\\
    &=&-\left[\frac{2\epsilon}{\mathcal{L}_X+3F+r}+\frac{\eta}{\mathcal{L}_X+3F+r}+\frac{r\beta}{\left(\mathcal{L}_X+3F+r\right)^2}+\frac{(6F-r)\epsilon}{\left(\mathcal{L}_X+3F+r\right)^2}\right]
    \nonumber\\&-&\frac{5}{3}\ln\sqrt{\frac{3(\mathcal{L}_X+3F+r)}{\mathcal{L}_X+3F}}\left[\frac{\mathcal{L}_X}{\mathcal{L}_X+3F}(c_s^{-2}-1)+2Z\frac{\mathcal{L}_Xc_s^{-2}+9F}{\mathcal{L}_X+3F}\right.
    \left.+\frac{\epsilon\varepsilon}{\left(\mathcal{L}_X+3F+r\right)^2}+\frac{6F\epsilon}{(\mathcal{L}_X+3F)(\mathcal{L}_X+3F+r)}\right]\nonumber\\
    &\simeq&-\frac{5}{3}\mathrm{ln}\sqrt{3(1+Q)}\left[\frac{\mathcal{L}_X}{\mathcal{L}_X+3F}(c_s^{-2}-1)+2Z\frac{\mathcal{L}_Xc_s^{-2}+9F}{\mathcal{L}_X+3F}\right],
\end{eqnarray}
\end{widetext}
where the first-order and second-order slow-roll small terms are ignored.
Recent observations by the PLANCK satellite place an upper bound of $|f_{NL}|\sim\mathcal{O}(10^2)$ \cite{PLANCKNG2018}, suggesting that the dominant factor in intrinsic non-Gaussianity, i.e., the sound speed $c_s$, should not be excessively small, and the NMDC characteristic quantity $F$ should not be excessively large. Meanwhile, the thermal dissipative strength may have a relatively large range, from the weak to the strong regime.

The non-Gaussian results obtained in NMDC noncanonical warm inflation can be consistent with those in GR noncanonical warm inflation \cite{Zhang2015} under the limit $F\rightarrow0$:
\begin{equation}\label{fnlintcanonical}
  |f_{NL}|\simeq\left(\frac{1}{c_s^2}-1\right)\ln\sqrt{\frac{3(\mathcal{L}_X+r)}{\mathcal{L}_X}}.
\end{equation}
These equations indicate that in GR noncanonical warm inflation, the non-Gaussian magnitude is slightly smaller than that in NMDC noncanonical warm inflation. The main difference is that in GR noncanonical warm inflation, the total nonlinear parameter $f_{NL}$ is dominated by both thermal and noncanonical effects, with the thermal contribution being more significant. In contrast, the presence of nonminimal gravitational coupling hinders the evolution of the inflaton field, resulting in a greater deviation from traditional potential-driven slow-roll inflation and a higher non-Gaussian level. Due to the competition between nonminimal coupling and thermal dissipation, the thermal contribution to the non-Gaussianity in NMDC noncanonical warm inflation is less pronounced. Under the limit $F\rightarrow0$ and $c_s\rightarrow1$, the general non-Gaussian results in NMDC noncanonical warm inflation can be consistently reduced to the canonical case, where $f_{NL}^{\delta N}=\frac{5\epsilon}{3(1+r)}-\frac{5\eta}{6(1+r)}-\frac{5r\epsilon}{6(1+r)^2}+\frac{5r\beta}{6(1+r)^2}$ \cite{Zhang2016}, $f_{NL}^{int}=-\frac56\ln\sqrt{3(1+r)}\frac{\epsilon\varepsilon}{(1+r)^2}$ \cite{Gupta2002,Gupta2006}. It is found that the term $f^{\delta N}_{NL}$ is a first-order slow-roll small quantity, while $f_{NL}^{int}$ is a second-order small quantity in canonical warm inflation. These equations suggest that the non-Gaussian feature in canonical warm inflation essentially differs from that in NMDC noncanonical warm inflation. In canonical warm inflation under the GR frame, the total nonlinear parameter $f_{NL}$ is dominated by the $f_{NL}^{\delta N}$ term instead of the $f_{NL}^{int}$ term.

\section{\label{sec4}conclusions and discussions}

In this paper, the total primordial non-Gaussianity generated in NMDC noncanonical warm inflation is investigated and analytically calculated. The framework for NMDC noncanonical warm inflation is presented, and some key equations of this scenario are given, such as the motion equation, e-folds, slow-roll equations, and the associated slow-roll conditions. Then, this paper focuses on the key issue: calculating the non-Gaussianity generated by NMDC noncanonical warm inflation. The nonlinear parameter, commonly employed to quantify the magnitude of non-Gaussianity, is divided into two parts: the $\delta N$ part $f^{\delta N}_{NL}$, and the intrinsic part $f_{NL}^{int}$, where the former describes the contribution from the four-point correlation of the dominant Gaussian perturbations of the inflaton field, while the latter comes from the three-point correlation, which represents the intrinsic non-Gaussianity of the inflaton field. These two parts together fully characterize the primordial non-Gaussianity in NMDC noncanonical warm inflation.

To calculate the $\delta N$-part non-Gaussianity, the $\delta N$ formalism is used. From the resulting expression, it is found that $f_{NL}^{\delta N}$ can be represented as a linear combination of the slow-roll parameters. This suggests that $f_{NL}^{\delta N}$ is a first-order slow-roll small quantity in this scenario, indicating that the $\delta N$-part non-Gaussianity in NMDC noncanonical warm inflation is not significant. This conclusion is consistent with the results from traditional GR-frame warm inflation. Nevertheless, the situation changes when calculating the intrinsic non-Gaussianity. The intrinsic non-Gaussianity comes from the three-point correlations of the inflaton field, and it mainly depends on the sound speed $c_s$, the NMDC characteristic quantities $F$, $Z$ and the thermal dissipation strength parameter $Q$. Our analysis of the total non-Gaussianity in NMDC noncanonical warm inflation reveals that $f_{NL}^{int}$ dominates over the $f_{NL}^{\delta N}$ part, and both the sound speed and NMDC effects play a key role in determining the total non-Gaussianity. Moreover, the presence of NMDC decreases the contribution of thermal effect compared to the case in GR-frame noncanonical warm inflation, suggesting that thermal and NMDC effects may compete in contributing to the non-Gaussianity. Thermal dissipation effects also contribute to the overall non-Gaussianity.

To summarize, although the general form of the intrinsic non-Gaussianity still exhibits the familiar $c_s^{-2}$ dependence and equilateral shape, its amplitude and physical origin are notably modified in the NMDC warm inflation scenario. The nonminimal derivative coupling fundamentally reshapes the non-Gaussian signature by introducing a gravitational friction that competes with thermal dissipation. This interplay not only partially suppresses the characteristic $c_s^{-2}$ enhancement known from noncanonical inflationary models but also generates a novel, comparable contribution governed by the coupling characteristic parameter $Z$. The result is a unique, hybrid non-Gaussianity whose dominant contribution transitions from being governed solely by the sound speed term to a regime of synergistic dominance by multiple parameters. These results underscore that the nontrivial combination of NMDC, noncanonical and thermal effects constitutes the main novel aspect of this work. Our future work will provide a detailed numerical calculation and observational constraints on some specific models of this scenario.

\section{Acknowledgments}
This work was supported by the Shandong Provincial Natural Science Foundation (Grant No. ZR2021MA037 and No. ZR2022JQ04), the National Natural Science Foundation of China (Grant No. 12575134 and No. 11605100)), the Natural Science Foundation of Henan (Grant No. 232300421351), and the Cultivation Project of Tuoxin Team in Henan University of Technology.

\end{document}